\def \tr {{\rm tr}}
\documentclass[twocolumn,showpacs,preprintnumbers,amsmath,amssymb]{revtex4}

\usepackage{epsf}

\begin{document}
\title{A symmetry-breaking mechanism for the Z4 general-covariant evolution
system}
\author{C.~Bona, T.~Ledvinka, C.~Palenzuela and M.~\v Z\' a\v cek
\\ Departament de Fisica, Universitat de les Illes Balears,
\\     Ctra de Valldemossa km 7.5, 07071 Palma de Mallorca, Spain}

\begin{abstract}
The general-covariant Z4 formalism is further analyzed. The gauge conditions are
generalized with a view to Numerical Relativity applications and the conditions
for obtaining strongly hyperbolic evolution systems are given both at the first
and the second order levels. A symmetry-breaking mechanism is proposed that
allows one, when applied in a partial way, to recover previously proposed
strongly hyperbolic formalisms, like the BSSN and the Bona-Mass\'o ones. When
applied in its full form, the symmetry breaking mechanism allows one to recover
the full five-parameter family of first order KST systems. Numerical codes based in the
proposed formalisms are tested. A robust stability test is provided by evolving
random noise data around Minkowski space-time. A strong field test is provided by
the collapse of a periodic background of plane gravitational waves, as described
by the Gowdy metric.
\end{abstract}

\maketitle
\section{Introduction}

The waveform emitted in the inspiral and merger of a relativistic binary is a
theoretical input crucial to the success of the laser interferometry gravitational
laboratories \cite{Ligo,Virgo,Geo,Tama}. Although the regular orbiting phase can be
treated with good accuracy by well-known analytical perturbation methods, the
later phases belong clearly to the strong field regime of either a black hole
or a neutron star collision, so that a computational approach is mandatory.
This kind of computational effort has been the objective of the Binary Black
Hole (BBH) Grand Challenge \cite{GC} and other world wide collaborations. The
resulting numerical codes are based on the so called ADM formalism \cite{ADM62}
for the Einstein field equations, where only a subset of the equations are
actually used for evolution whereas the remaining ones are considered as constraints to
be imposed on the initial data only (free evolution approach \cite{Cent80}).

It is clear that, by taking the constraints out of the evolution system, one is extending
the solution space. This extension is the crucial step that opened the way to
new hyperbolic formalisms after the seminal work of Y. Choquet-Bruhat and
T. Ruggeri \cite{CR83}, using the constraints to modify the evolution
system in many different ways \cite{BM92,BM95,FR94,FR96,SN95,BS99,AY99,Hern00,KST01,ST02},
even taking additional derivatives \cite{CR83,AACY95,Frie96}. These formalisms can be
interpreted as providing many non-equivalent ways of extending the solution space
of Einstein's equations with at least one common feature: constraint equations
are left out of the final evolution system.

This means that the resulting systems do have an extended solution space, which includes
constraint-violating solutions in addition to the ones verifying the original Einstein's
equations. As far as the constraints are first integrals of the extended evolution system,
Einstein's solutions could be computed by solving the constraints equations only for the initial
data (free evolution). But, unless some enforcing mechanism is used during the subsequent
time evolution, numerical errors will activate constraint-violating modes. Numerical
simulations can deal with such modes, at least when the deviations from Einstein's solutions
are moderate. But it happens that large deviations are usually associated with instabilities
of constraint-violating modes, leading to code crashing.

In particular, numerical codes based on these new hyperbolic formalisms happen to be quite
intolerant to violations of the Hamiltonian constraint in the initial data.
This is a serious drawback if one is planning to use the results of the
analytical approximation to the regular orbiting phase of a binary system as
initial data for a numerical simulation of the final ringdown and merger. The
numerical code will crash before any template for the gravitational wave emission
could be extracted. This is because the analytical data are just a good
approximation, so that the energy constraint does not hold exactly and the code
is intolerant to that kind of ``off-shell" initial data. Although there can be other
options, we claim that a numerical code tolerant to constraint violation would be
undoubtedly the best alternative.

There have been some attempts in that direction. In Ref.
\cite{BFHR99} the technique of Lagrange multipliers was used for
including the constraints into the dynamical system. The extended
system includes the Lagrange multipliers as additional dynamical
fields ($\lambda$-system). A further step along that direction is
given in Ref. \cite{SY}, where the extended system is "adjusted"
by further combining the evolution equations with the constraints,
including then many extra arbitrary parameters. In both cases, the
goal is to enforce the constraints in a dynamical way. One can
monitor the errors by looking at the "subsidiary system", which
can be derived from Bianchi identities and can be interpreted as
the evolution system for constraint deviations. Parameters are
adjusted in a way such that the characteristic speeds of the
subsidiary system are either real and non-zero (so that the
corresponding errors propagate away to the boundaries) or they
have the right sign in the imaginary part to enforce damping
(instead of exploding) the constraint deviations.

A simpler option (but not the only one, see for instance
\cite{GGKM03}) for including the constraints into the evolution system 
would be the general-covariant extension of the Einstein field equations proposed
recently (Z4 system) \cite{Z4}:
\begin{equation}\label{EinsteinZ4}
  R_{\mu \nu} + \nabla_{\mu} Z_{\nu} + \nabla_{\nu} Z_{\mu} =
  8\; \pi\; (T_{\mu \nu} - \frac{1}{2}\, T\; g_{\mu \nu})~,
\end{equation}
so that the full set of dynamical fields {consists of} the pair
$\{g_{\mu\nu},Z_\mu\}$. The solutions of the original field
equations can then be recovered by imposing the algebraic constraint
\begin{equation}\label{Zis0}
  Z_{\mu} = 0
\end{equation}
and the evolution of this constraint is subject to the
linear homogeneous equation
\begin{equation}\label{WaveZis0}
  \Box\; Z_{\mu} + R_{\mu\nu} Z^\nu = 0~,
\end{equation}
which can be easily obtained from (\ref{EinsteinZ4}) allowing for
the contracted Bianchi identities. Here again, by allowing for non-zero values of
the extra four-vector $Z_\mu$, one is extending the solution space. But now, as we
will see in the following section, one is using all the field equations to evolve
the pair $\{ g_{\mu\nu}, Z_\mu \}$: no equation is taken out of the system and, as
a result, general covariance is not broken. The initial metric $g_{\mu\nu}$ can be
taken to be the one arising from analytical approximations and the initial
four-vector $Z_\mu$ can be taken to vanish without any kind of inconsistence: one
can even use the evolving values of $Z_\mu$ during the calculation as a good covariant
indicator of the quality of the approximation.

Notice that (\ref{WaveZis0}) plays here the role of the subsidiary
system. It is adjusted \textit{ab initio}, without any parameter
fine-tuning, because light speed is the only characteristic speed
in (\ref{WaveZis0}). Constraint deviations, that is non-vanishing
values of $Z^\mu$ will then propagate to the boundaries. The fact
that our constraints (\ref{Zis0}) are algebraic will greatly
simplify the task of providing outgoing boundary conditions that
let the constraint deviations get out of the numerical grid
\cite{BeA}

Besides these considerations, there are other important
theoretical issues that we will address in this work. The first
one is a thorough analysis of the hyperbolicity of the evolution
system. This is more or less straightforward for the first order
version of the system, as discussed in Appendix B, but it is not
so well known in the case of the second order version, discussed
in Appendix A, where we have used the results of Kreiss and Ortiz
\cite{KO01} that recently shed light on this issue, which is
crucial to discuss the well-posedness \cite{KL89} of the evolution
system (see also \cite{Reu03} for similar results for the BSSN
system).

The second theoretical point that we want to stress here is that the Z4 system
(\ref{EinsteinZ4}) is not just one more hyperbolic formalism to be added to the
long list. As far as it is the only general covariant one, the question arises
whether the existing non-covariant hyperbolic formalisms \cite{BM92,BM95,FR94,FR96,SN95,BS99,AY99,Hern00,KST01,ST02}
can be recovered from (\ref{EinsteinZ4}) by some ``symmetry breaking" mechanism. We
have extended in this sense a previous work \cite{Z3} where the deep relationship
between the more widely used hyperbolic formalisms was pointed out. A partial
symmetry breaking mechanism is presented in section II for recovering the second
order systems \cite{SN95,BS99} and for the first order systems
containing additional dynamical fields \cite{BM92,BM95} in section III. These
sections are followed by another one containing numerical simulations that have
been proposed recently \cite{Mexico} as standard test-beds for Numerical Relativity codes.
A more general symmetry breaking mechanism is proposed in Appendix C to recover
first order formalisms which do not contain additional dynamical fields
\cite{FR94,FR96,AY99,Hern00,KST01,ST02}.

\section{3+1 Evolution Systems}

The general-covariant equations (\ref{EinsteinZ4}) can be written
in the equivalent 3+1 form \cite{Z4} (Z4~evolution system)
\begin{eqnarray}
\label{dtgamma}
  (\partial_t -{\cal L}_{\beta})~ \gamma_{ij}
  &=& - {2\;\alpha}\;K_{ij}
\\
\label{dtK}
   (\partial_t - {\cal L}_{\beta})~K_{ij} &=& -\nabla_i\alpha_j
    + \alpha\;   [{}^{(3)}\!R_{ij}
    + \nabla_i Z_j+\nabla_j Z_i
\nonumber \\
    &-& 2K^2_{ij}+(\tr K-2\Theta)\;K_{ij}
\nonumber \\
    &-& S_{ij}+\frac{1}{2}\,(\tr\; S -\; \tau)\;\gamma_{ij}\;]
\\
\label{dtTheta}
(\partial_t -{\cal L}_{\beta})~\Theta &=& \frac{\alpha}{2}\;
 [{}^{(3)}\!R + 2\; \nabla_k Z^k + (\tr K - 2\; \Theta)\;\tr K
\nonumber \\
&-&  \tr(K^2)  - 2\; Z^k {\alpha}_k/\alpha - 2\tau]
\\
\label{dtZ}
 (\partial_t -{\cal L}_{\beta})~Z_i &=& \alpha\; [\nabla_j\;({K_i}^j
  -{\delta_i}^j \tr K) + \partial_i \Theta
\nonumber \\
  &-&2\; {K_i}^j\; Z_j  -  \Theta\, {{\alpha}_i/ \alpha} - S_i]
\end{eqnarray}
where we have noted
\begin{equation}\label{tauSdef}
  \Theta \equiv  \alpha \; Z^0,~
  \tau \equiv  8 \pi  \alpha^2\; T^{00},~
  S_i \equiv  8 \pi \alpha \; T^0_{\;i},~
  S_{ij} \equiv 8 \pi \;T_{ij}.
\end{equation}

In the form (\ref{dtgamma}-\ref{dtZ}), it is evident that the Z4 evolution system
consists only of evolution equations. The only constraints (\ref{Zis0}), that can
be translated into:
\begin{equation}\label{ZThis0}
  \Theta~=~0,\qquad Z_i~=~0,
\end{equation}
are algebraic so that the full set of field equations (\ref{EinsteinZ4}) is
actually used during evolution. This is in contrast with the ADM evolution system
\cite{ADM62}, which can be recovered from (\ref{dtgamma}-\ref{dtZ}) by imposing
(\ref{ZThis0}). The first two equations (\ref{dtgamma},\ref{dtK}) would transform
into the well known ADM evolution system, whereas the
last two equations (\ref{dtTheta},\ref{dtZ}) would transform into the
standard energy and momentum constraints, namely
\begin{eqnarray}
 {}^{(3)}\!R + \tr^2 K -  {\tr}(K^2)  &=& 2\; \tau
\label{constraint1} \\
 \nabla_j\;({K_i}^j -{\delta_i}^j trK) &=& S_i
\label{constraint2}
\end{eqnarray}

In the ``free evolution" ADM approach \cite{Cent80}, both (\ref{constraint1}) and
(\ref{constraint2}) were taken out of the evolution system: they were imposed only
on the initial data. This was consistent because
(\ref{constraint1},\ref{constraint2}) are first integrals of the ADM evolution
system, but one can not avoid violations of (\ref{constraint1},\ref{constraint2})
due to errors in numerical simulations or approximated initial data, as stated
before. As a result, numerical simulations will deal as well with extended solutions.
The main difference with the Z4 case, aside from covariance considerations, is that
in the Z4 case the quantities $Z_\mu$ describing constraint deviations are included
into the evolution system.

One can also ask in this context what happens if, instead of
imposing of the full set (\ref{ZThis0}), one imposes the single
condition
\begin{equation}\label{This0}
  \Theta~=~0
\end{equation}
obtaining a system with only three supplementary dynamical
variables $Z_i$ of the kind determined in \cite{Z3} (Z3 system):
the one corresponding to the parameter choice
\begin{equation}\label{mu2nun0}
  \mu~=~2,~~~ \nu~=~n~=~0
\end{equation}
(we follow the notation of \cite{Z3}).

One can easily understand two of the three conditions (\ref{mu2nun0}),
namely
\begin{equation}\label{mu2nuisn}
  \mu~=~2,~~~ \nu~=~n,
\end{equation}
because this amounts to the `physical speed' requirement for the
degrees of freedom not related to the gauge \cite{Z3} and nothing
else can arise from the general covariant equations
(\ref{EinsteinZ4}) which are at our starting point. But values of
the remaining parameter $n$ other than zero would be very
interesting. In particular, the choice
\begin{equation}\label{mu2nun43}
  \mu~=~2,~~~ \nu~=~n~=~\frac{4}{3}.
\end{equation}
would lead to the evolution system which is quasiequivalent
(equivalent principal parts \cite{Z3}) to the well known BSSN
system \cite{SN95,BS99}.

At this point, let us consider the following recombination of the dynamical fields
\begin{equation}\label{Ktilde}
  \tilde{K}_{ij}~\equiv~K_{ij} - \frac{n}{2}\; \Theta\; \gamma_{ij}
\end{equation}
so that the Z4 system (\ref{dtgamma}-\ref{dtZ}) can be written in
a one-parameter family of equivalent forms just by replacing
everywhere
\begin{equation}\label{Ktildeinv}
  {K_{ij}}~\rightarrow~\tilde{K}_{ij} + \frac{n}{2}\; \Theta\; \gamma_{ij}.
\end{equation}
This kind of transformations leave invariant the solution space of
the system (it is actually the same system expressed in a
different set of independent fields). But if the suppression of
the $\Theta$ field (\ref{This0}) is made after the replacement
(\ref{Ktildeinv}), one gets a one-parameter family of
non-equivalent systems (Z3 evolution systems), namely:
\begin{eqnarray}
  (\partial_t -{\cal L}_{\beta})\; \gamma_{ij}
  &=& - {2\;\alpha}\;K_{ij}
\label{Z3dtgamma} \\
\nonumber
   (\partial_t -{\cal L}_{\beta}) K_{ij} &=& -\nabla_i\alpha_j
    + \alpha\;   [{}^{(3)}\!R_{ij}
    + \nabla_i Z_j + \nabla_j Z_i
\nonumber \\
    &-& 2K^2_{ij}+\tr K\;K_{ij}
 - S_{ij}+\frac{1}{2}(\tr S - \tau)\gamma_{ij}]
\nonumber \\
 &-& \frac{n}{4} \; \alpha\; [{}^{(3)}\!R + 2\; \nabla\!\! \cdot\! Z
 + \tr^2 K - \tr(K^2)
\nonumber \\
 &&\;\;\;\;\;\;\;\; - 2 (\alpha^{-1} {\alpha_k }) Z^k - 2 \tau]\;\gamma_{ij}
\label{Z3dtK} \\
\label{Z3dtZ}
 (\partial_t -{\cal L}_{\beta}) Z_i &=& \alpha\; [\nabla_j\;({K_i}^j
  -{\delta_i}^j~ \tr K) -2 {K_i}^j Z_j - S_i]
\nonumber \\
\end{eqnarray}
where we have suppressed the tilde over $K_{ij}$, allowing for
the vanishing of $\Theta$.

The resulting system (\ref{Z3dtgamma}-\ref{Z3dtZ}) is
quasiequivalent (equivalent principal parts) to the `system A' in
ref. \cite{Z3} verifying the `physical speed' requirement
(\ref{mu2nuisn}). As we have already mentioned, it follows that
the particular case
\begin{equation}\label{nis43}
  n ~=~ \frac{4}{3}
\end{equation}
is quasiequivalent to the BSSN system \cite{SN95,BS99}. The system (\ref{Z3dtgamma}-\ref{nis43}) can be decomposed into trace and trace-free
parts
\begin{equation}\label{conformal_metric}
   e^{4\;\phi} ={\gamma}^{1/3} \;,\quad
  {\tilde{\gamma}}_{ij} = e^{-4\;\phi}\;\gamma_{ij}
\end{equation}
\begin{equation}\label{conformal_curvature}
  K = \gamma^{ij}\;K_{ij} \;,\quad
  {\tilde{A}}_{ij} = e^{-4\;\phi}\;(K_{ij}-
  \frac{1}{3}\; K\;\gamma_{ij})
\end{equation}
\begin{equation}\label{Gs2}
  {\tilde{\Gamma}}_i = -{\tilde{\gamma}}_{ik}\;{{\tilde{\gamma}}^{kj}}_{\;,j}
                       + 2\; Z_i
\end{equation}
to follow the correspondence with BSSN more closely. It must be pointed out,
however, that one does not get in this way the original BSSN system: there is
actually one difference in the lower order terms (only the principal parts are
equivalent). The difference is in the term of the form
\begin{equation}\label{Z3AZterm}
+\frac{n}{2} \alpha_k\; Z^k\; \gamma_{ij}
\end{equation}
in the evolution equation (\ref{Z3dtK}), which is missing in the original
BSSN system \cite{BS99}. This lower order term is needed for consistency with the general
covariant equations (\ref{EinsteinZ4}).

We have seen then how the widely used ADM and BSSN systems can be obtained
from the more general Z4 formalism. The equivalence transformation
(\ref{Ktilde}) plays the crucial role because suppressing the
$\Theta$ field (\ref{This0}) produces a sort of symmetry breaking:
different values of the parameter $n$ will lead to evolution
systems that can no longer be transformed one into another once
the set of dynamical fields is reduced by the disappearance of
$\Theta$. It can be regarded as a partial symmetry breaking mechanism for the
original equations (\ref{dtgamma}-\ref{dtTheta}). The terms ``partial'' refers to the
fact that only the quantity $\Theta$ is suppressed, while the $Z_i$ are kept into the
system (\ref{Z3dtgamma}-\ref{Z3dtZ}). A complete symmetry breaking mechanism is
discussed in Appendix C.

In section IV, we present the results of some test-bed
simulations for the ADM and Z4 systems. We have considered for
simplicity only vacuum space-times with the time coordinate
conditions
\begin{equation}\label{dtAlpha}
 (\partial_t -{\cal L}_{\beta})~\ln \alpha = -~\alpha\;[f {\tr} K - \lambda \Theta]
\end{equation}
which are a further generalization of the one proposed in
\cite{Z4}, where
\begin{equation}\label{f1lambda2}
  f~=~1,~~~ \lambda~=~2.
\end{equation}
This two-parameter family of coordinate conditions is very
interesting from the point of view of Numerical Relativity
applications. But it is also interesting from the theoretical
point of view, because it provides the opportunity to apply
the recent results of Kreiss and Ortiz \cite{KO01} on the hyperbolicity
of the ADM system in a wider context. In Appendix A, we will use
the same formulation (see ref. \cite{KL89} for more details)
to study the hyperbolicity of the Z4 system
with the two-parameter family of dynamical gauge conditions
(\ref{dtAlpha}).

\section{First Order Systems}

A first order version of the Z4 evolution system
(\ref{dtgamma}-\ref{dtZ}) can be obtained in the standard way by
considering the first space derivatives
\begin{equation}\label{AkDkij}
 A_k~\equiv~\alpha_k/\alpha,~~D_{kij}~\equiv~\frac{1}{2}\;\partial_k \gamma_{ij}
\end{equation}
as independent dynamical quantities with evolution equations
given by
\begin{equation}\label{dtA}
 \partial_t A_k~+~\partial_k [~ \alpha (f\; {\tr}\; K - \lambda \Theta) ~]~=~0
\end{equation}
\begin{equation}\label{dtD}
 \partial_t D_{kij}~+~\partial_k [ \alpha K_{ij} ]~=~0
\end{equation}
(we will consider in what follows the vanishing shift case for
simplicity), so that the full set of dynamical fields can be given
by
\begin{equation}\label{uvector}
 u ~ = ~ \{\alpha,~\gamma_{ij},~ K_{ij},~ A_k,~D_{kij},~\Theta,~Z_k \}
\end{equation}
(38 independent fields).

Care must be taken when expressing the Ricci tensor
${}^{(3)}\!R_{ij}$ in (\ref{dtK}) in terms of the derivatives
of $D_{kij}$, because as far as the constraints
(\ref{AkDkij}) are no longer enforced, the identity
\begin{equation}\label{dDdD}
 \partial_r D_{sij} = \partial_s D_{rij}
\end{equation}
can not be taken for granted in first order systems. As a
consequence of this ordering ambiguity of second derivatives, the
principal part of the evolution equation (\ref{dtK}) can be
written in a one-parameter family of non-equivalent ways, namely
\begin{equation}\label{PrincipalK}
 \partial_t K_{ij} ~+~\partial_k~[\alpha~\lambda^k_{ij}]~=~...~
\end{equation}
\begin{eqnarray}\label{deflambda}
 \lambda^k_{ij} &\equiv& -{\Gamma}^k_{~ij}+
 \frac{1 - \zeta}{2}\, (D_{ij}^{~~k}+D_{ji}^{~~k}
   -\delta^k_i D_{rj}^{~~r}-\delta^k_j D_{ri}^{~~r})
\nonumber \\
  &+& {\frac{1}{2}}\, \delta^k_i(A_j+{D_{jr}}^r-2Z_j)+
  {\frac{1}{2}}\, \delta^k_j(A_i+{D_{ir}}^r-2Z_i)
\nonumber \\
\end{eqnarray}
so that the parameter choice $\zeta = +1$ corresponds
to the standard Ricci decomposition
\begin{equation}\label{Def3R}
{}^{(3)}\!R_{ij}~=~\partial_k\;{\Gamma^k}_{ij}-\partial_i\;{\Gamma^k}_{kj}
+{\Gamma^r}_{rk}{\Gamma^k}_{ij}-{\Gamma^k}_{ri}{\Gamma^r}_{kj}
\end{equation}
whereas the opposite choice $\zeta = -1$ corresponds to the
de~Donder-Fock \cite{DeDo21,Fock59} decomposition
\begin{eqnarray}\label{Def3dDF}
{}^{(3)}\!R_{ij}&=&-\partial_k\;{D^k}_{ij}+\partial_{(i}\;{\Gamma_{j)k}}^{k}
- 2 {D_r}^{rk} D_{kij} \nonumber \\
&+& 4 {D^{rs}}_i D_{rsj} - {\Gamma_{irs}} {\Gamma_j}^{rs}-{\Gamma_{rij}} {\Gamma^{rk}}_k
\end{eqnarray}
which is most commonly used in Numerical Relativity codes.  Note
that this ambiguity does not affect to the principal part of eq.
(\ref{dtTheta}), namely
\begin{equation}
 \partial_t\Theta + \partial_k [\alpha V^k] = ...
\end{equation}
where we have noted
\begin{equation}\label{defVk}
 V_k \equiv {D_{kr}}^r - {D^r}_{rk} -Z_k.
\end{equation}

We are now in position to discuss the hyperbolicity of this first
order version of the Z4 systems. This is done in a straightforward
way in Appendix A.

In order to compare the new first order system with the Bona-Mass\'o ones
\cite{BM92,BM95} we could either apply here again the
recombination (\ref{Ktilde}) followed by the suppression
(\ref{This0}) of the $\Theta$ field or we could take directly a
first order version of the Z3 system
(\ref{Z3dtgamma}-\ref{Z3dtZ}).
Eqs. (\ref{Z3dtgamma},\ref{Z3dtZ},\ref{dtD})  do not change, but eqs.
(\ref{dtA},\ref{PrincipalK},\ref{deflambda}) are
replaced in this case by
\begin{eqnarray}
\label{PPZ3dtA}
  \partial_t A_k +\partial_k~[\alpha~f~\tr K]&=&0 \\
\label{PPZ3dtK}
 \partial_t K_{ij} ~+~\partial_k~[\alpha~{\lambda^k}_{ij}]&=& \ldots
\end{eqnarray}
\begin{eqnarray}\label{PPZ3deflambda}
 \lambda^k_{ij} &\equiv& -{\Gamma}^k_{~ij}
 -\frac{n}{2}\,V^k\gamma_{ij}
\\
 &+& \frac{1}{2} \delta^k_i(A_j+{D_{jr}}^r-2Z_j)
 + \frac{1}{2} \delta^k_j(A_i+{D_{ir}}^r-2Z_i)
\nonumber \\
  &+& \frac{1-\zeta}{2} (D_{ij}^{~~k}+D_{ji}^{~~k}
   -\delta^k_i D_{rj}^{~~r}-\delta^k_j D_{ri}^{~~r})
\nonumber
\end{eqnarray}
The full Bona-Masso family of evolution equations is recovered for
the $\zeta=-1$ case, where (\ref{PPZ3deflambda}) can be written as
\begin{eqnarray}\label{PPBMdeflambda}
 \lambda^k_{ij} &\equiv& {D}^k_{~ij}-\frac{n}{2} V^k\gamma_{ij}
\\
 &+&\frac{1}{2} \delta^k_i(A_j-{D_{jr}}^r+2V_j)+
 \frac{1}{2} \delta^k_j(A_i-{D_{ir}}^r+2V_i)
\nonumber
\end{eqnarray}
with $V_k$ defined by (\ref{defVk}).

In the following section, we will compare the behavior of both
families in numerical simulations. To this end, we will also
consider the first order version of the ADM system, which can be
obtained from the previous ones just by suppressing the $Z_i$
eigenfields.
\begin{equation}\label{Z3is0}
  Z_i~=~0~.
\end{equation}

Before proceeding to the test section, let us just mention, that
the same game of recombining the $\Theta$ field with $K_{ij}$
(\ref{Ktildeinv}) before suppressing it can also be played with
the $Z_k$ fields and $D_{kij}$ in first order systems. As stated before,
this will provide a complete symmetry breaking mechanism. We will do
that in Appendix C, where we will show how the well known KST system
\cite{KST01} can also be recovered in that way from
the Z4 framework discussed in this paper.

\section{Testing Second and First Order Systems}

We will present in this section a couple of numerical experiments
which have been suggested very recently \cite{Mexico} as standard
test-beds for Numerical Relativity codes.  Our philosophy is that
all the tests could be done ``out of the box", using well known
numerical methods and the equations that are fully presented here:
anyone should be able to reproduce our results without recourse to
additional information.

We will use the standard method of lines \cite{MoL} as a finite differencing
algorithm, so that space and time discretization will be dealt separately.
Space differencing will consist of taking centered
discretizations of derivatives in our 3D grid. We use the standard centered
stencil for first derivatives and we make sure that second derivatives,
when needed, are coded also as centered derivatives of these first
derivatives, even if it takes up to five point along every axis.In order to
avoid boundary effects, the grid has the topology of a three-torus, with
periodic boundaries along every axis. The time evolution will be
dealt with a third order Runge-Kutta algorithm. The time step ${\rm d}t$ is kept
small enough to avoid an excess of numerical dissipation that could distort
our results in long runs.

\subsection{Robust stability test}

Let us consider a small perturbation of Minkowski space-time which
is generated by taking random initial data for every dynamical
field in the system. The level of the random noise must be small
enough to make sure that we will keep in the linear regime even
for a thousand of crossing times (the time that a light ray will
take to cross the longest way along the numerical domain). This is
in keeping with the theoretical framework of Appendix A, where
only linear perturbations around the Minkowski metric are
considered.

\begin{figure}[t]
\begin{center}
\epsfxsize=8cm
\epsfbox{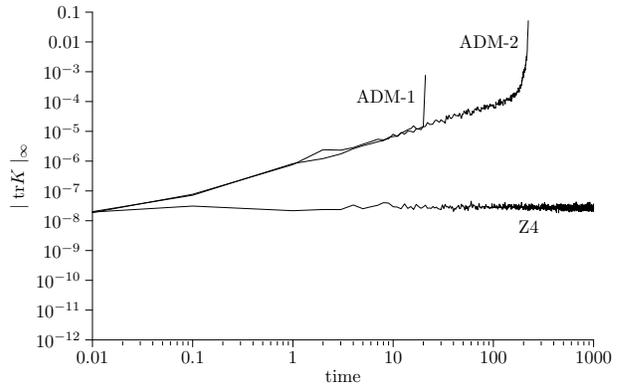}
\end{center}
\caption{\label{plot1} The maximum of (the absolute value of) $trK$ is
plotted against the number of crossing times in a logarithmic scale. The
initial level of random noise remains constant during the evolution in the
case of strongly hyperbolic systems (only the second order Z4 system is shown
here for clarity). In the case of weakly hyperbolic systems, like the ADM
second order system ADM-2 or its first order version ADM-1, a linear growth
is detected up to the point where the codes crash. Notice that the second
order version is more robust, an order of magnitude, than the first one.
The simulations are made with 50 grid points with ${\rm d}t=0.03\;{\rm d}x$.}
\end{figure}

We have plotted in Fig.~\ref{plot1} our results for the standard harmonic
case (\ref{f1lambda2}). We see the expected polynomial (linear in this case)
growth \cite{KL89} of the weakly hyperbolic ADM system.
Notice that modifications of the lower order terms (the ones not
contributing to the principal part) could lead to catastrophic
exponential growth, revealing an ill-posed evolution system
\cite{KL89}. In this paper, however, we will limit ourselves to discussing the linear
regime as an hyperbolicity test for the principal part of the system. In this sense,
the linear growth of the ADM plots in Fig.~\ref{plot1} confirms
the weakly hyperbolic character of the ADM system.

\begin{figure}[t]
\begin{center}
\epsfxsize=8cm
\epsfbox{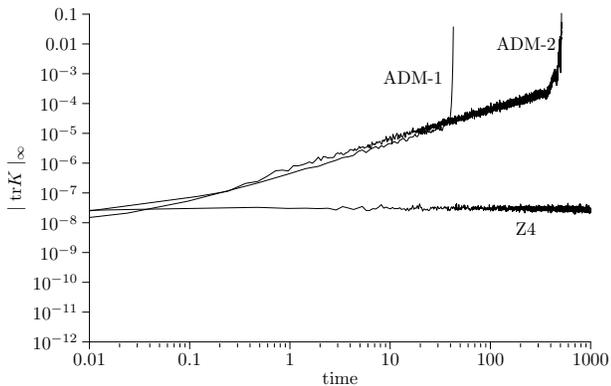}
\end{center}
\caption{\label{plot1bis} Same as Fig.~\ref{plot1}, but with less time resolution
(${\rm d}t=0.06\;{\rm d}x$ with the same ${\rm d}x$). There is a slight amount of
dissipation that delays the crashing of the ADM codes; this is especially visible
for the second order version ADM-2, which keeps being more robust than its first order
counterpart ADM-1. The behaviour of the Z4 code keeps unaffected.}
\end{figure}

The Z4 system shows instead the no-growth behavior, independent
of the time resolution (see Figs.~\ref{plot1},~\ref{plot1bis}), which one would expect from
a strongly hyperbolic system. The same qualitative behavior is shown by the
corresponding Z3 systems in (\ref{Z3dtgamma}-\ref{Z3dtZ}),
including the one with $n=4/3$ that is quasiequivalent to the BSSN
system.

\begin{figure}[b]
\caption{\label{plot3} Array of results of numerical experiments
in the gauge parameter plane ($f$,$\lambda$), by using the Z4
system. A triangle stands for linear growth of noise (weak
hyperbolicity), whereas a cross stands for a constant noise level
(strong hyperbolicity). This is consistent with the strong
hyperbolicity requirements predicted in Appendix B: either $f =1$
and $\lambda=2$, or $f \neq 1$ and $f> 0$.}
\begin{center}
\epsfbox{fig3.ps}
\end{center}
\end{figure}

\begin{figure}[t]
\begin{center}
\epsfxsize=8cm
\epsfbox{fig2.ps}
\end{center}
\caption{\label{plot2} Same as Fig.~\ref{plot1}, but using the
second order ICN method to evolve in time instead of a third order
Runge-Kutta algorithm. Numerical dissipation is severely
distorting the plots, by masking the linear growth in the weakly
hyperbolic case and dramatically reducing the initial noise level
in the strongly hyperbolic case. Notice than both ${\rm d}t$ and
${\rm d}x$ are the same as in Fig.~\ref{plot1} and we are using
also the same space discretization algorithm: only the time
evolution method has changed.}
\end{figure}

We also show in Fig.~\ref{plot2} the same results, but distorted by using
too much numerical dissipation: the time evolution here is dealt
with the second order ICN method rather than the third order
Runge-Kutta of Fig.~\ref{plot1}. After hundreds of crossing times, the
numerical dissipation manages to curve the linear growing of ADM
and the noise level goes down in the Z4 case. This is just a
numerical artifact, because in the linear regime there is no
physical damping mechanism for strongly hyperbolic systems in a
three-torus, where periodic boundary conditions do not allow
propagation outside the domain. This is why we will use here third
order Runge-Kutta instead of the ICN method proposed in
\cite{Mexico}.

In Fig.~\ref{plot3}, we explore parameter space in the ($f$,$\lambda$)
plane. If we interpret the constant behavior in Fig.~\ref{plot1} as
revealing a strongly hyperbolic system and the polynomial growth
(linear in this case) in Fig.~\ref{plot1} as revealing a weakly hyperbolic system, the
results of our numerical experiment fully agree with the
theoretical results presented in Appendix A.

\subsection{Gowdy waves}

In order to test the strong field regime, let us consider now the Gowdy solution
\cite{Gowdy71}, which describes a space-time containing plane polarized gravitational
waves (see also \cite{Berger} for an excellent review of these space-times as
cosmological models). The line element can be written as
\begin{equation}\label{gowdy_line}
  {\rm d}s^2 = t^{-1/2}\, e^{Q/2}\,(-{\rm d}t^2 + {\rm d}z^2)
  + t\,(e^P\, {\rm d}x^2 + e^{-P}\, {\rm d}y^2)
\end{equation}
where the quantities $Q$ and $P$ are functions of $t$ and $z$ only
and periodic in $z$, so that (\ref{gowdy_line}) is well suited for
finite difference numerical grids with periodic boundary
conditions along every axis. Following \cite{Mexico}, we will
choose the particular case
\begin{eqnarray}
\label{function_P}
  P &=& J_0 (2 \pi t)\; \cos(2 \pi z)
\\
\label{fucntion_L}
  Q &=&  \pi J_0 (2 \pi) J_1 (2 \pi)
   -2 \pi t J_0 (2 \pi t) J_1 (2 \pi t) \cos^2(2 \pi z)
\nonumber \\
  &+& 2 \pi^2 t^2 [{J_0}^2 (2 \pi t)  + {J_1}^2 (2 \pi t)
  - {J_0}^2 (2 \pi)  - {J_1}^2 (2 \pi)]
\nonumber \\
\end{eqnarray}
so that it is clear that the lapse function
\begin{equation}\label{gowdy_lapse}
  \alpha = t^{-1/4}\; e^{Q/4}
\end{equation}
is constant everywhere at any time $t_0$ at which $J_0(2\pi t_0)$ vanishes. In \cite{Mexico}
the initial slice $t=t_0$ was chosen for the simulation of the collapse,
where $2 \pi t_0$ is the 20-th root of the Bessel function $J_0$, i.e. $t_0\; \simeq \;9.88$.

\begin{figure}[t]
\begin{center}
\epsfxsize=8cm 
\epsfbox{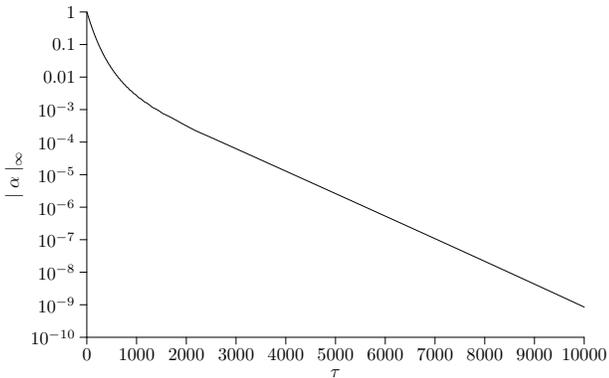}
\end{center}
\caption{\label{plot4} Time evolution of (the maximum value of) the lapse function
$\alpha$ in a collapsing Gowdy space-time (harmonic slicing). Notice that the harmonic
time coordinate $\tau$ is not the proper time and it does not coincide with the
number of crossing times, due to the collapse of the lapse, which is visible here, by a
$10^{-9}$ factor.}
\end{figure}

Let us now perform the following time coordinate transformation
\begin{equation}\label{gowdy_time}
  t~=~t_0\;e^{-\tau / \tau_0}, ~~ \tau_0~=~t_0^{3/4} e^{Q(t_0)/4}~\simeq~472\;,
\end{equation}
so that the expanding line element (\ref{gowdy_line}) is seen in the new time
coordinate $\tau$ as collapsing towards the $t=0$ singularity, which is approached
only in the limit $\tau\rightarrow\infty$. This ``singularity avoidance" property of the $\tau$
coordinate is not surprising if one realizes that the resulting slicing by $\tau=constant$
surfaces is harmonic \cite{BM83}.

\begin{figure}[b]
\caption{\label{plot5} The quantity $\Theta$ is plotted as an
indicator of the accumulated error of the simulations for the ADM,
Z4 and Z3-BSSN second order forms. Even in this logarithmic scale,
it can be clearly seen how the Z4 and Z3-BSSN codes perform much
better than the ADM one: error differs by one order of magnitude
at $\tau \simeq 1000$. The Z3-BSSN code gets closer to the Z4 one
in the oscillatory phase (up to $\tau \simeq 2000$).}
\begin{center}
\epsfxsize=8cm 
\epsfbox{fig-g2.ps}
\end{center}
\end{figure}

This means that we can launch our simulations starting with a
constant lapse $\alpha_0=1$ at $\tau=0$ ($t=t_0$) with the gauge
parameter choice $f=1$ (which means also $\lambda=2$ in the Z4
case). Notice that the harmonic time coordinate $\tau$ is not the
proper time and it does not coincide with the number of crossing
times, due to the collapse of the lapse. Remember also that the
local value of light speed (proper distance over coordinate time)
is $\pm \alpha$. Even though in our plots $\tau$ goes up to
$10000$, the light ray manages to cross the domain in the
$z$-direction only $t_0 \simeq 9.88$ times, as it follows from the
original form (\ref{gowdy_line}) of the line element.

\begin{figure}[t]
\begin{center}
\epsfxsize=8cm 
\epsfbox{fig8.ps}
\end{center}
\caption{\label{plot6bis} Convergence test for the Z4 code. The three lines
correspond to $50$, $100$ and $200$ grid points. The quantity $\Theta$ itself is a
direct measure of the error. In that logarithmic scale differences of $log\, 4$
correspond to dividing by four the error when doubling the resolution. This second
order convergence rate is clearly shown in the figure.}
\end{figure}

\begin{figure}[b]
\caption{\label{plot6} Same as Fig.~\ref{plot5}, but with the
first order versions of both the ADM and Z4 codes, which behave in
the same way as their second order counterparts. The Z3-BM code
gets here closer the ADM one in the oscillatory phase (up to $\tau
\simeq 2000$), in contrast with the behavior of the Z3-BSSN code
in Fig.~\ref{plot5}.}
\begin{center}
\epsfxsize=8cm 
\epsfbox{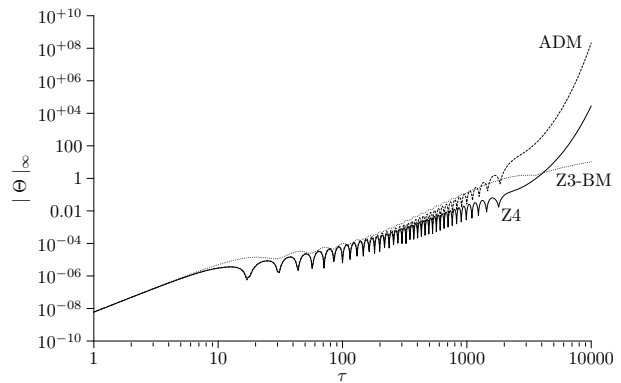}
\end{center}
\end{figure}

We plot in Fig.~\ref{plot4} the maximum values of the lapse
function as time goes on, measured in terms of the harmonic time
coordinate $\tau$. Notice the huge magnitude of the dynamical
space we are covering, as $\alpha$ goes down (collapse of the
lapse) by the factor of one billion during the simulation. This is
a  real challenge for numerical codes and all of them are doing
quite well until $\tau=1000$. The behavior at later times is
dominated by the lower order terms: coordinate light speed ($\pm
\alpha$) is so small that the dynamics of the principal part is
frozen and care must be taken to avoid too big time steps. We have
not seen any of the codes crashing even in very long simulations
(up to $\tau = 60.000$) when the size of the time step is kept
under control.

In Fig.~\ref{plot5}, we use the quantity $\Theta$, as computed
from (\ref{dtTheta}) to monitor the quality of the simulation both
in the Z4 case (\ref{dtgamma}-\ref{dtZ}), where $\Theta$ is a
dynamical field, and in the other two second order cases (ADM and
Z3-BSSN), where $\Theta$ is no longer a dynamical quantity but can
still be used as a good measure of the error in the simulation.
This makes easier to perform convergence tests (second order convergence
is shown in Fig.~\ref{plot6bis}). Notice that our Z3-BSSN code performs
here much better than the original BSSN one \cite{BS99}, as reported in
\cite{Mexico}. This is mainly due to the fact that we are not using here the conformal
decomposition (\ref{conformal_metric}-\ref{Gs2}) which is at odds
with the structure of the line element (\ref{gowdy_line}). This is
why we talk here about Z3-BSSN (\ref{Z3dtgamma}-\ref{nis43})
instead of simply BSSN.

\begin{figure}[b]
\caption{\label{plot7} Same as Fig.~\ref{plot5}, but with
different values of the parameter $n$ arising from the symmetry
breaking mechanism in the Z3 codes. The differences show up in the
collapse final phase (starting at $\tau \simeq 2000$). Notice that
the value $n=4/3$ corresponds to the Z3-BSSN case in
Fig.~\ref{plot5}, whereas the case $n=1$ corresponds to (the
second order version of) the Z3-BM case in Fig.~\ref{plot6}. First
order versions (not shown) behave in the same way as their second
order counterparts shown here}
\begin{center}
\epsfxsize=8.38cm 
\epsfbox{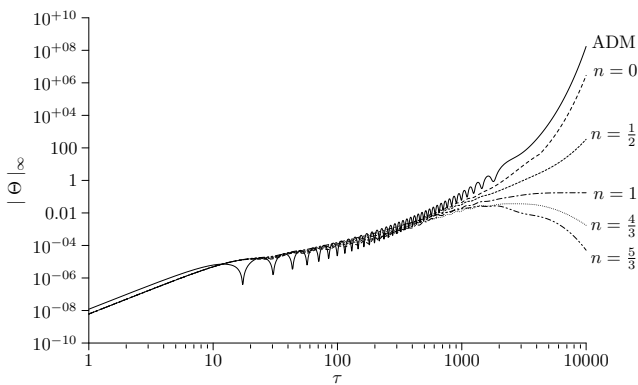}
\end{center}
\end{figure}

The same kind of comparison is made in Fig.~\ref{plot6} for the
first order version of the ADM and Z4 codes, which show the same
behavior than their second order counterparts in Fig.~\ref{plot5}.
The third plot corresponds to the Z3 version of the Bona-Mass\'o
code, which can be obtained from
(\ref{PPZ3dtA}-\ref{PPZ3deflambda}), with the parameter choice
$\zeta = -1$, $n=+1$. Notice that in the oscillatory phase (up to
$\tau \simeq 2000$) the Z3-BM code tends to behave like the ADM
one, whereas in Fig.~\ref{plot5} the Z3-BSSN code tends to behave
more like the Z4 one.

Notice that the Z3 versions, when compared with both ADM and Z4,
show a different behavior after the oscillatory phase: $\theta$
grows at a much lower rate or even starts going down. This
behavior is due to the extra terms that appear in the evolution
equations for $K_{ij}$ after the symmetry breaking, which can be
controlled by the parameter $n$ introduced in (\ref{Ktildeinv}).
This point is clearly shown in Fig.~\ref{plot7}, where different
choices of $n$ produce different behavior in the final collapse
phase (starting at $\tau \simeq 2000$). This shows the relevance
of the recombination between the dynamical fields in numerical
applications, as pointed out in \cite{KST01}. Further details and other
numerical tests can be found in the webpage at \textit{http://stat.uib.es}.

\renewcommand{\theequation}{A.\arabic{equation}}
\setcounter{equation}{0}
\section*{Appendix A: Hyperbolicity of the Second Order Z4 System}

Let us consider the linearized version of the Z4 system
(\ref{dtgamma}-\ref{dtZ}) around Minkowski space-time in order to
study the propagation of a plane wave in that background:
\begin{eqnarray}
  \gamma_{ij} &=& \delta_{ij} + 2\; e^{i\;
  \omega \cdot x}\; {\hat{\gamma}}_{ij} (\omega,t)
\label{pert_gamma} \\
  \alpha &=& 1 +  e^{i\;\omega \cdot x}\; \hat{\alpha}
(\omega,t)
\label{pert_alpha} \\
  K_{ij} &=& i\; \omega\; e^{i\;\omega \cdot x}\;
{\hat{K}}_{ij} (\omega,t)
\label{pert_K} \\
  \Theta &=& i\; \omega\; e^{i\;\omega \cdot x}\;
{\hat{\Theta}} (\omega,t)
\label{pert_Theta} \\
  Z_{k} &=& i\; \omega\; e^{i\;\omega \cdot x}\;
{\hat{Z}}_{k} (\omega,t)
\label{pert_Z}
\end{eqnarray}
where we will take for simplicity $\beta^i = 0$ and
\begin{equation}\label{shortcuts}
  \omega_k = \omega\;n_k \:,\qquad \delta^{ij}\; n_i\; n_j = 1
\end{equation}

The Z4 system reads then
\begin{eqnarray}
\partial_t {\hat{\gamma}}_{ij} &=& - i\; \omega\; {\hat{K}}_{ij}
\label{A_dtgamma} \\
\partial_t {\hat{\alpha}} &=& - i\; \omega\; [f\;tr{\hat{K}} -
\lambda\;{\hat{\Theta}}]
\label{A_dtalpha} \\
\partial_t {\hat{\Theta}} &=& - i\; \omega\; [tr{\hat{\gamma}} -
{\hat{\gamma}}^{nn}-{\hat{Z}}^n]
\label{A_dtTheta} \\
\partial_t {\hat{Z}}_k &=& - i\; \omega\; [n_k\; (tr{\hat{K}} -
{\hat{\Theta}}) - {{\hat{K}}_k}^n]
\label{A_dtZ} \\
\partial_t {\hat{K}}_{ij} &=& - i\; \omega\; {\hat{\lambda}}_{ij}
\label{A_dtK}
\end{eqnarray}
where we have noted
\begin{eqnarray}\label{A_deflambda}
{\hat{\lambda}}_{ij} &\equiv& {\hat{\gamma}}_{ij} +
n_i\;n_j\;({\hat{\alpha}} + tr{\hat{\gamma}})
\nonumber \\
&-& n_i\;({{\hat{\gamma}}_j}^n + {\hat{Z}}_j) -
n_j\;({{\hat{\gamma}}_i}^n + {\hat{Z}}_i)
\end{eqnarray}
and where the symbol $n$ replacing an index means the contraction
with $n_i$. It can be also expressed in matrix form, namely
\begin{eqnarray}\label{2Z4-linear-matrix}
  {\hat{u}} &=& ({\hat{\alpha}},{\hat{\gamma}}_{ij},{\hat{K}}_{ij},
  {\hat{\Theta}},{\hat{Z}}_k) \\
  \partial_t {\hat{u}} &=& - i\; \omega\; {\bf A}\; {\hat{u}}
\end{eqnarray}

The spectral analysis of the characteristic matrix ${\bf A}$ provides
the following list of eigenvalues and eigenfields

\renewcommand{\labelenumi}{\alph{enumi})}
\begin{enumerate}
\item Standing eigenfields (zero characteristic speed)
\begin{equation}\label{A_EF0}
  {\hat{\alpha}} - f\/ \tr{\hat{\gamma}} + \lambda (\tr{\hat{\gamma}} -
  {\hat{\gamma}}^{nn} - {\hat{Z}}^n) \,,\quad
  {{\hat{\gamma}}^{n}}_{\perp} + {\hat{Z}}_{\perp}
\end{equation}
where the symbol $\perp$ replacing an index means the projection orthogonal
to $n_i$.
\item Light-cone eigenfields (characteristic speed $\pm 1$)
\begin{eqnarray}
 && {\hat{K}}_{\perp \perp} \pm {\hat{\gamma}}_{\perp \perp}
\label{A_EFLtt} \\
 && {{\hat{K}}_{\perp}}^n \pm {\hat{Z}}_{\perp}
\label{A_EFLnt} \\
 && {\hat{\Theta}} \pm [tr{\hat{\gamma}} - {\hat{\gamma}}^{nn} -
 {\hat{Z}}^n ] \label{A_EFLen}\;.
\end{eqnarray}
Notice that (\ref{A_EFLtt}-\ref{A_EFLen}) can
be seen as the components of a tensor:
\begin{eqnarray}\label{A_EFLij}
  {\hat{L}}_{ij}^{\pm} &\equiv& [{\hat{K}}_{ij} - (tr{\hat{K}}-2{\hat{\Theta}})\;
  n_i\;n_j]
\nonumber \\
  &\pm& [{\hat{\lambda}}_{ij} - {\hat{\alpha}}\;n_i\;n_j]
\end{eqnarray}

\item Gauge eigenfields (characteristic speed $\pm \sqrt{f}$)
\begin{eqnarray}\label{A_EFG}
  {\hat{G}}^{\pm} &\equiv& \sqrt{f} \left[\tr{\hat{K}} + \frac{2-\lambda}{f-1}\;{\hat{\Theta}}\right]
\nonumber \\
   &\pm& \left[{\hat{\alpha}} + \frac{2\;f-\lambda}{f-1}(\tr{\hat{\gamma}}
    - {\hat{\gamma}}^{nn} - {\hat{Z}}^n )\right]
\end{eqnarray}
\end{enumerate}

From (\ref{A_EF0}-\ref{A_EFG}) we can easily conclude \cite{KO01,KL89}
\renewcommand{\labelenumi}{\roman{enumi})}
\begin{enumerate}
\item All the characteristic speeds are real (weak hyperbolicity at least) if and
only if $f\geq 0$.

\item In the case $f=0$, the two components of the gauge pair
(\ref{A_EFG}) are not independent, so that the total number of
independent eigenfields is 16 instead of 17 required for strong
hyperbolicity.

\item The case $f=1$ (harmonic case) is special:
\begin{itemize}
\item if $\lambda \ne 2$, then the gauge pair (\ref{A_EFG}), which
can be previously rescaled by a $(f-1)$ factor, is equivalent to
(\ref{A_EFLen}), so that one has only 15 independent eigenfields
\item if $\lambda=2$, then the quotient $\frac{2-\lambda}{f-1}$
can take any value, reflecting the degeneracy of the gauge and
light cone eigenfields. One can then recover the full set of 17
independent eigenfields (strong hyperbolicity).
\end{itemize}
\item In all the remaining cases ($f>0$, $f \ne 1$), the system is
strongly hyperbolic, as we can recover the full set of 17
independent eigenfields.
\end{enumerate}

\renewcommand{\theequation}{B.\arabic{equation}}
\setcounter{equation}{0}
\section*{Appendix B: Hyperbolicity of the First Order Z4 System}

The principal part of the first order Z4 evolution system
(\ref{dtgamma}-\ref{dtZ},\ref{dtAlpha},\ref{dtA}-\ref{dtD})
can be written as (vanishing shift case):
\begin{eqnarray}
\label{B_dtgamma}
&\!\!\!\!\!& \partial_t \gamma_{ij} = \ldots \;,\qquad  \partial_t \alpha = \ldots
\\
\label{B_dtTheta}
&\!\!\!\!\!& \partial_t \Theta +\partial_k~[\alpha~V^k] = \ldots
\\
\label{B_dtZ} &\!\!\!\!\!& \partial_t Z_i +
\partial_k~[\alpha~(\delta^k_i(\tr K-\Theta)-K^k_{~i})] =\ldots
\\
\label{B_dtA}
&\!\!\!\!\!& \partial_t A_k +\partial_k~[\alpha~(f\/ \tr K - \lambda \Theta)]=\ldots
\\
\label{B_dtD}
&\!\!\!\!\!& \partial_t D_{kij} +\partial_k~[\alpha~K_{ij}] =\ldots
\\
\label{B_dtK}
&\!\!\!\!\!& \partial_t K_{ij} +\partial_k~[\alpha~\lambda^k_{ij}] =\ldots
\end{eqnarray}
where
\begin{eqnarray}\label{B_deflambda}
 \lambda^k_{ij} &=& {D^k}_{ij}
   -{\frac{1+\zeta}{2}} (D_{ij}^{~~k}+D_{ji}^{~~k}
   -\delta^k_i D_{rj}^{~~r}-\delta^k_j D_{ri}^{~~r})
\nonumber \\
 &+& \frac{1}{2}\, \delta^k_i(A_j-{D_{jr}}^r+2V_j)
\nonumber \\
 &+& \frac{1}{2}\, \delta^k_j(A_i-{D_{ir}}^r+2V_i)
\\
\label{B_defV}
  V_k &\equiv& {D_{kr}}^r-{D_{rk}}^{r}-Z_k.
\end{eqnarray}

Now, if we consider the propagation of perturbations with
wavefront surfaces given by the unit (normal) vector $n_i$, we can
express (\ref{B_dtgamma}-\ref{B_dtK}) in matrix form
\begin{equation}\label{B_dtu}
  \alpha^{-1} \partial_t~u ~+~ {\bf A}(u) ~n^k~\partial_k u = ...~,
\end{equation}
where
\begin{equation}\label{B_uvector}
 u ~ = ~ \{\alpha,~\gamma_{ij},~ K_{ij},~ A_k,~D_{kij},~\Theta,~Z_k \}~.
\end{equation}
(notice that derivatives tangent to the wavefront surface play no
role here).

A straightforward analysis of the characteristic matrix ${\bf A}(u)$
provides the following list of eigenfields:

\renewcommand{\labelenumi}{\alph{enumi})}
\begin{enumerate}
\item Standing eigenfields (zero characteristic speed)
\begin{equation}\label{B_EF0}
 \alpha,~ \gamma_{ij},~A_\perp,~D_{\perp ij},~A_k-f D_k+\lambda V_k
\end{equation}
(24 independent fields), where the symbol $\perp$ replacing an
index means the projection orthogonal to $n_i$:
\begin{equation}\label{B_DAij}
 D_{\perp ij} \equiv D_{kij} - n_k n^r D_{rij}.
\end{equation}

\item Light-cone eigenfields (characteristic speed $\pm 1$)
\begin{eqnarray}\label{B_EFLij}
 {L^{\pm}}_{ij} &\equiv& [K_{ij}-n_i n_j \;{\tr}K ]
\nonumber \\
  &\pm& [{\lambda^n}_{ij} - n_i n_j~{\tr}\;\lambda^n]
\\
\label{B_EFL}
  L^{\pm} &\equiv& \theta \pm V^n
\end{eqnarray}
(12 independent fields), where the symbol $n$ replacing the index
means the contraction with $n_i$
\begin{equation}
 \lambda^n_{ij}~\equiv~n_k~\lambda^n_{ij}.
\end{equation}

\item Gauge eigenfields (characteristic speed $\pm \sqrt{f}$)
\begin{eqnarray}\label{B_EFG}
 G^{\pm} &\equiv& \sqrt{f} \left[{\tr}~K + \frac{2-\lambda}{f-1}\,\Theta \right]
\nonumber \\
  &\pm& \left[A^n + \frac{2f-\lambda}{f-1}\,V^n \right]
\end{eqnarray}
\end{enumerate}

From (\ref{B_EF0}-\ref{B_EFG}) we can easily conclude that

\renewcommand{\labelenumi}{\roman{enumi})}
\begin{enumerate}
\item All the characteristic speeds are real (weak hyperbolicity
at least) if and only if $f\geq 0$.

\item In the case $f=0$, the two components of the pair
(\ref{B_EFG}) are not independent, so that the total number of
independent eigenfields is 37 instead of 38 required for strong
hyperbolicity.

\item The case f=1 (harmonic case) is special:
\begin{itemize}
\item if $\lambda \ne 2$, then the pair of fields (\ref{B_EFG}) is
the same as (\ref{B_EFL}), so that one has only 36 independent
eigenfields \item if $\lambda=2$, then the quotient
$\frac{2-\lambda}{f-1}$ can take any value due to the degeneracy
of the gauge and light eigenfields. One can then recover the full
set of 38 independent eigenfields (strong hyperbolicity).
\end{itemize}
\item The first order Z4 system described by
(\ref{B_dtgamma}-\ref{B_dtK}) is strongly hyperbolic in all the
remaining cases ($f>0$, $f \ne 1$).
\end{enumerate}

Notice also that the special (harmonic) case $f=1$, $\lambda=2$
has been shown in \cite{Z4} to be symmetric hyperbolic for the
parameter choice $\zeta=-1$. The corresponding energy function
can be written as
\begin{eqnarray}\label{B_Energy}
  E &\equiv&
 K^{ij}K_{ij} + \lambda^{kij}\lambda_{kij} +
  (\tr{K} - 2\Theta)^2 + A^k A_k
\nonumber \\
  &+& (A^k - {D^{kr}}_r + 2 V^k)(A_k - {D_{kr}}^r + 2 V_k)
\end{eqnarray}
but notice that this expression is far from being unique. For
instance, allowing for (\ref{B_EF0}), the last term in
(\ref{B_Energy}) could appear with any arbitrary factor.

\renewcommand{\theequation}{C.\arabic{equation}}
\setcounter{equation}{0}
\section*{Appendix C: Recovering the KST systems}

Let us start with the first order Z4 evolution system
(\ref{dtgamma}-\ref{dtZ},\ref{dtAlpha},\ref{dtA}-\ref{dtD}) where
the principal part is given by (\ref{B_dtgamma}-\ref{B_defV}). Now
let us follow the two step `symmetry breaking' process, namely

\begin{enumerate}
\item Recombine the dynamical fields $K_{ij}$, $D_{kij}$ with
$\Theta$ and $Z_i$ in a linear way,
\begin{eqnarray}
\label{C_Ktilde}
  \tilde{K}_{ij}&=&K_{ij} - {\frac{n}{2}}\; \Theta\; \gamma_{ij}~,
\\
\label{C_dtilde}
  d_{kij}&=& 2 D_{kij} + \eta\; \gamma_{k(i} Z_{j)} + \chi\; Z_k \gamma_{ij}~,
\end{eqnarray}
where we have used the notation of Ref. \cite{KST01}, replacing
only their parameter $\gamma$ by $-n/2$ for consistency. Notice that
(\ref{C_Ktilde}-\ref{C_dtilde}) is generic in the sense that it is the most
general linear combination that preserves the tensor character of the dynamical fields
under linear coordinate transformations.

\item Suppress both $\theta$ and $Z_i$ as dynamical fields, namely
\begin{equation}\label{C_ZThis0}
  \Theta~=~0,\qquad Z_i~=~0~.
\end{equation}
\end{enumerate}

In that way, the principal part (\ref{B_dtgamma}-\ref{B_defV})
becomes
\begin{eqnarray}\label{C_dtgamma}
 \partial_t \gamma_{ij} &=& \ldots \;, \qquad \partial_t \alpha = \ldots
\\
\label{C_dtA}
 \partial_t A_k &+&\partial_k~[\alpha~f\;\tr\; \tilde{K}\; ]~=~0
\\
\label{C_dtd}
   \partial_t d_{kij} &+& \partial_r [
    \alpha \{2\; \delta^r_k\; \tilde{K}_{ij}
    -\chi\; ({\tilde{K}_{k}}^r-\delta^r_k\; \tr \tilde{K}) \gamma_{ij}
\nonumber \\
   &+& \eta\; \gamma_{k(i}({{\tilde{K}}^r}_{~j)}   - {\delta^r}_{j)} \tr \tilde{K})\}] =
   \ldots
\\
\label{C_dtK}
 \partial_t \tilde{K}_{ij} &+&\partial_k~[\alpha~\lambda^k_{ij}]=\ldots
\end{eqnarray}
\begin{eqnarray}\label{C_deflambda}
   2\; {\lambda^k}_{ij} &=& {d^k}_{ij} - \frac{n}{4}({d_{kr}}^r-{{d_r}^{rk}})\gamma_{ij}
\nonumber \\
   &+& \frac{1+\zeta}{2}({d_{ij}}^k + {d_{ji}}^k )
   - \frac{1-\zeta}{2}( \delta_i^k\, {d_{rj}}^r+ \delta_j^k\, {d_{ri}}^r)
\nonumber \\
   &+& \delta^k_j\, (A_i+\frac{1}{2} {d_{ir}}^r)+ \delta^k_i\, (A_j+\frac{1}{2} {d_{jr}}^r)
\end{eqnarray}

for the reduced set of variables
\begin{equation}\label{C_uvector}
 u ~ = ~ \{\alpha,~\gamma_{ij},~ \tilde{K}_{ij},~ A_k,~d_{kij} \}.
\end{equation}

This provides a ``dynamical lapse" version \cite{ST02} of the KST
evolution systems. In order to recover the original ``densitized
lapse" version, one must in addition integrate explicitly  the
dynamical relationship (\ref{dtAlpha}) between the lapse and the
volume element (remember that now $\Theta=0$). It can be easily
done in the case
\begin{equation}\label{C_fdef}
 f ~=~2\;\sigma~=~{\rm constant},
\end{equation}
namely
\begin{equation}\label{C_dtAlpha}
 \partial_t(\alpha \gamma^{-\sigma} ) ~=~ 0,
\end{equation}
so that the value of $\alpha$ can be defined in terms of $\gamma$ for every initial
condition. The same thing can be done with $A_i$ and $d_i$, so that
\begin{equation}\label{C_Adef}
 A_i ~\equiv~\sigma\;{d_{ir}}^r~+~...
\end{equation}
and the set of dynamical fields is then further reduced to
\begin{equation}\label{C_KSTuvector}
 u ~ = ~ \{\gamma_{ij},~ K_{ij},~d_{kij} \}.
\end{equation}
The principal part of the evolution system is then given by (we
suppress the tildes over the $K_{ij}$)

\begin{eqnarray}
\label{BKSTdtgamma}
 \partial_t \gamma_{ij} &=& \ldots
\\
\label{BKSTdtd}
 \partial_t d_{kij} &+& \partial_r [
  \; \alpha \{\,2\, \delta^r_k\, {K}_{ij} -\chi\;({{K}_{k}}^r-\delta^r_k\, \tr {K}) \gamma_{ij}
\nonumber \\
  &+&  \eta\;\gamma_{k(i}({{K}^r}_{j)}   - \delta^r_{j)} \tr {K})\, \}\;] = \ldots
\\
\label{BKSTdtK}
 \partial_t K_{ij} &+& \partial_k~[\alpha~\lambda^k_{ij}] = \ldots
\end{eqnarray}
\begin{eqnarray}\label{BKSTdeflambda}
   2\; {\lambda^k}_{ij} &=& {d^k}_{ij}
   - \frac{n}{4}({d_{kr}}^r-{{d_r}^{rk}})\gamma_{ij}
\nonumber \\
   &-& \frac{1-\zeta}{2}(\delta^k_i\, {{d^r}_{rj}}+ \delta^k_j \, {{d^r}_{ri}})
   + \frac{1+\zeta}{2}({d_{ij}}^k + {d_{ji}}^k )
\nonumber \\
   &+& \frac{1+2\sigma}{2}( \delta^k_i\, {d_{jr}}^r+ \delta^k_j\, {d_{ir}}^r)
\end{eqnarray}

which corresponds precisely to (the principal part of) the original KST system \cite{KST01}.

{\em Acknowledgements: This work has been supported by the EU Programme
'Improving the Human Research Potential and the Socio-Economic
Knowledge Base' (Research Training Network Contract HPRN-CT-2000-00137),
by the Spanish Ministerio de Ciencia y Tecnologia through the research
grant number BFM2001-0988 and by a grant from the Conselleria d'Innovacio
i Energia of the Govern de les Illes Balears.}

\bibliographystyle{prsty}

\end{document}